\documentstyle[11pt]{article}
\def\theequation{1.\arabic{equation}}
\normalbaselines
\begin{document}
\bibliographystyle{unsrt}

%\large
\title{\LARGE CORRELATIONS IN PARTICLE PRODUCTION \\
AND SQUEEZING PHENOMENA}
\author{\Large  I.M. Dremin\\
                   Lebedev Physical Institute,\\
                   Leninsky Prospekt, 53, 117924, Moscow, Russia}
\date{}
\maketitle
\begin{abstract}
Recent developments in studies of multiparticle correlations in high energy
particle collisions are reviewed. Both experimental data and theoretical
results in quantum chromodynamics are discussed. Application of the developed
methods to the coherent, squeezed and correlated states of photons is
considered. Some speculations concerning possible coherent and
squeezed states of pion fields as well as their specific features
 are described.

\end{abstract}
\setcounter{equation}{0}
\begin{center}
{\Large 1. INTRODUCTION}
\end{center}
The study of correlations of quantum fields is a common subject in
statistical physics, quantum optics, multiparticle production. Therefore,
it is not at all surprising to find out common methods of research in
those fields. I shall describe some newly obtained results in studies
of correlations of particles produced in high energy collisions as well as
their relation to the correlations of photons and, in particular, to the
squeezing phenomenon.

The nature of any source of radiation (of photons, gluons or other
entities) can be studied by analyzing multiplicity distributions,
energy spectra, various correlation properties, etc. A particular
example is provided by the coherent states of fields which give rise to
Poisson distribution. However, in most cases one has to deal with various
distributions revealing different dynamics. In particle physics dealing with
strong interactions, one relies
on quantum chromodynamics (QCD) as a substitute of quantum electrodynamics
(QED) in case of photons. The fields of quarks and gluons are considered
instead of electrons and photons. Let us stress that quarks are described by
spinors similar to electrons, and gluons are massless vector particles like
photons. The important difference is that gluons carry the color charge while
the photons are electrically neutral. It gives rise to the self-interaction
of gluons as well as to famous properties of the asymptotic freedom and
confinement.

Let us turn now to some experimental facts about high energy particle
interactions. When two high energy particles collide, the bunch of new
particles is produced in each event. It is common to plot the distribution of
those events in the number of particles produced (the multiplicity
distribution, for short). Usually, they differ drastically from the Poisson
distribution, clearly indicating strong correlations. Several years ago the
experimentalists of UA5 Collaboration in CERN noticed \cite{UA5} a shoulder in
the multiplicity distribution of particles produced in $p\bar p$ collisions at
energies ranging from 200 to 900 GeV in the center of mass system. (Let me
remind here that the mass of the proton is about 1 GeV, i.e. much smaller.)
It looked like a small wiggle over a smooth curve and was immediately ascribed
by theorists to processes with larger number of
Pomerons exchanged in the traditional schemes. More recently, several
collaborations studying $e^{+}e^{-}$ collisions at 91 GeV in CERN  reported
(see, e.g., \cite{DELPHI,OPAL}) that they failed to fit the multiplicity
distributions of produced particles by smooth curves (the Poisson and Negative
Binomial distributions were among them). Moreover, subtracting such smooth
curves from the experimental ones they found steady oscillatory behaviour of
the difference. It was ascribed to the processes with different number of jets.

 The new sensitive method of theoretical analysis of distributions
was proposed in \cite{Dremin,DreminPhysLett} (for the review
see \cite{DreminUFN}).
It appeared as a byproduct of the solution of the equations for generating
functions of multiplicity distributions in quantum chromodynamics (QCD).
It appeals to the moments of the multiplicity distribution. According to QCD,
the so--called cumulant moments (or just cumulants), described in more
details below, should reveal the oscillations as functions of their ranks,
while they are identically equal to zero for the Poisson distribution
and are steadily decreasing positive
functions for Negative Binomial distribution so widely used in
phenomenological analysis. Experimental data show the oscillatory behaviour
of cumulants (see \cite{Gian}) of multiplicity distributions in high energy
inelastic processes initiated by various particles and nuclei,
even though some care should be taken due to
the high multiplicity cut-off of the data. When applied to the squeezed states,
the method demonstrates \cite{Pavel} the oscillations of cumulants in slightly
squeezed states and, therefore, can be useful for their detection.
In such circumstances one is tempted to speculate about the alternative
explanation when considering
possible similarity of these findings to typical features of squeezed and
correlated states.
However, first let us describe briefly the method of analysis of the
distributions which we rely upon.
\setcounter{equation}{0}
\def\theequation{2.\arabic{equation}}

\begin{center}
{\Large 2. DISTRIBUTION FUNCTION AND ITS MOMENTS.}
\end{center}

Here I summarize definitions and
notations for the values that are used to characterize various processes
of inelastic scattering according to number of particles produced in these
processes. The relations connecting these values to
each other and the examples related to some distributions typical in the
probability theory are provided.

Any process of inelastic scattering (that is the scattering with new
particles produced) can be characterized by the function $P_{n}$, the
multiplicity distribution function. The value of $P_{n}$ denotes the
probability to observe $n$ particles produced in the collision. It
is clear that $P_{n}$ must be normalized to unity:
\begin{eqnarray}
  \sum_{n=0}^{\infty} P_{n}=1.
\end{eqnarray}
Sometimes the multiplicity distribution of particles produced can be
conveniently described by its moments. It means that the series of numbers
$P_{n}$ is replaced by another series of numbers according to a certain rule.
All these moments can be obtained by the differentiation of the so--called
generating function $G(z)$ defined by the formula:
\begin{eqnarray}
  G(z)=\sum_{n=0}^{\infty} P_{n} z^{n}.
\end{eqnarray}
Thus, instead of the discrete set of numbers $P_{n}$ we can study the
analytical function $G(z)$.

We will use the factorial moments and cumulants defined by the following
relations:
\begin{eqnarray}
    &&F_{q}=\frac {\sum_{n=0}^{\infty }P_{n}(n-1)\cdots (n-q+1)}
    {\left(\sum_{n=0}^{\infty }P_{n}n\right)^{q}}
    =\left.\frac{1}{\langle n \rangle^{q}}\frac{d^{q}G(z)}{dz^{q}}
    \right|_{z=1},
\end{eqnarray}
\begin{eqnarray}
  \left.K_{q}=\frac{1}{\langle n \rangle^{q}}\frac{d^{q}\mbox{ln}G(z)}
  {dz^{q}} \right|_{z=1},
\end{eqnarray}
where
\begin{eqnarray}
  \langle n \rangle =\sum_{n=0}^{\infty }P_{n}n
\end{eqnarray}
is the average multiplicity.

Reciprocal formulas expressing the generating function in terms of cumulants
and factorial moments can also be obtained:
\begin{eqnarray}
  \lefteqn{}
    &&G(z)=\sum_{q=0}^{\infty} \frac{z^{q}}{q!} \langle n \rangle^{q} F_{q}
    \nonumber\\
    &&(F_{0}=F_{1}=1),
\end{eqnarray}
\begin{eqnarray}
  \lefteqn{}
    &&\mbox{ln}G(z)=\sum_{q=1}^{\infty} \frac{z^{q}}{q!} \langle n
    \rangle^{q} K_{q} \nonumber\\
    &&(K_{1}=1).
\end{eqnarray}
The probability distribution function itself is related to the generating
function in the following way:
\begin{eqnarray}
  \left. P_{n}=\frac{1}{n!}\frac{d^{n}G(z)}{dz^{n}} \right|_{z=0}.
\end{eqnarray}
Factorial moments and cumulants are connected to each other by the following
recursion relation:
\begin{eqnarray}
  F_{q}=\sum_{m=0}^{q-1}C_{q-1}^{m} K_{q-m} F_{m},
\end{eqnarray}
where
\begin{eqnarray}
  C_{q-1}^{m} = \frac{(q-1)!}{m! (q-m-1)!}
  \nonumber
\end{eqnarray}
are binomial coefficients.
Relation (2.9) gives the opportunity to find the factorial moments if
cumulants are known, and vice versa.

It must be pointed out that cumulants are very sensitive to small variations
of the distribution function and hence can be used to distinguish the
distributions which otherwise look quite similar.

Usually, cumulants and factorial moments for the distributions occuring in
the particle physics are very fast growing with increase of their rank.
Therefore sometimes it is more convenient to use their ratio:
\begin{eqnarray}
  H_{q}=\frac{K_{q}}{F_{q}} ,
\end{eqnarray}
which behaves more quietly with increase of the rank $q$ remaining a
sensitive measure of the tiny details of the distributions.

In what follows we imply that the rank of the distribution function moment is
non--negative integer even though formulas (2.3) and (2.4) can be generalized
to the non--integer ranks.

Let us demonstrate two typical examples of the distribution.

$1.\hspace{0.5cm} Poisson \hspace{0.4cm} distribution.$

The Poisson distribution has the form:
\begin{eqnarray}
  P_{n}=\frac{\langle n \rangle^{n}}{n!}\mbox{exp}\left(-\langle n \rangle
  \right).
\end{eqnarray}
Generating function (2.2) can be easily calculated:
\begin{eqnarray}
  G(z)=\mbox{exp}\left(\langle n \rangle z\right).
\end{eqnarray}
According to (2.3) and (2.4) we have for the moments of this distribution:
\begin{eqnarray}
  F_{q}=1, \hspace{0.5cm} K_{q}=H_{q}=\delta_{q1}.
\end{eqnarray}

$2.\hspace{0.5cm}Negative\hspace{0.4cm}binomial\hspace{0.4cm}distribution.$

This distribution is rather successfully used for fits of main features of
experimental data in particle physics. It has the form:
\begin{eqnarray}
  P_{n}=\frac{\Gamma(n+k)}{\Gamma(n+1)\Gamma(k)}\left(\frac{\langle n \rangle}
  {k}\right)^{n}\left(1+\frac{\langle n \rangle}{k}\right)^{-(n+k)},
\end{eqnarray}
where $\Gamma$ is the gamma function, and $k$ is a fitting parameter.

At $k=1$ we have the usual Bose distribution. The Poisson distribution can be
obtained from (2.15) in the limit $k\rightarrow \infty$.

Generating function for the negative binomial distribution reads:
\begin{eqnarray}
  G(z)=\left( 1-\frac{z\langle n \rangle}{k} \right)^{-k},
\end{eqnarray}
and the moments of this distribution are:
\begin{eqnarray}
  \lefteqn{}
    &&F_{q}=\frac{\Gamma(k+q)}{\Gamma(k)k^{q}}, \nonumber\\
    &&K_{q}=\frac{\Gamma(q)}{k^{q-1}}, \nonumber\\
    &&H_{q}=\frac{\Gamma(q)\Gamma(k+1)}{\Gamma(k+q)}.
\end{eqnarray}
\setcounter{equation}{0}
\def\theequation{3.\arabic{equation}}

\begin{center}
{\Large 3. QCD EQUATIONS AND MOMENTS OF DISTRIBUTIONS.}
\end{center}

The multiparticle production processes are described in quantum chromodynamics
as a result of the interaction of quarks and gluons which leads to creation
of additional quarks and gluons forming the observed hadrons at the very last
stage. The most typical features of the processes are determined by the vector
nature of gluons and by the dimensionless coupling constant. The gluons are
colour charged in distinction to photons which have no electric charge.
Therefore,
they can emit gluons in addition to quark-antiquark pairs. That is why both
quark and gluon jets are considered in quantum chromodynamics as main
objectives.
Their development is described by the evolution equations. The main parameter
of
the evolution is the opening angle of the jet or its transverse momentum. The
subsequent emission of gluons and quarks fills in the internal regions of the
previously developed angular cones so that they do not overlap (angular
ordering). This
remarkable property
can be exploited to formulate the probabilistic scheme for the development of
the jet as a whole. Then its evolution equations remind the well-known
classical
Markovian equations for the "birth--death" (or "mother--daughter") processes.
(The detailed discussion of that approach, based on the coherence phenomenon,
see in \cite{5}).

It is quite natural to start our studies with the simplest case of
gluodynamics.
There are no quarks in that case, and interactions of gluons are considered
only.
The system of equations degenerates to the single
equation
\begin{equation}
G^{\prime }(y) = \int _{0}^{1}dxK(x)\gamma _{0}^{2}[G(y+\ln x)G(y+\ln (1-x)) -
G(y)] ,   \label{56}
\end{equation}
where  $G^{\prime }(y)=dG/dy$, $y$ is the evolution parameter,
\begin{equation}
\gamma _{0}^{2} =\frac {6\alpha _S}{\pi } ,                \label{52}
\end{equation}
$\alpha _S$ replaces $\alpha $ of electrodynamics and the kernel of the
equation is
\begin{equation}
K(x) = \frac {1}{x} - (1-x)[2-x(1-x)] .    \label{53}
\end{equation}
 It is the non-linear integro-differential equation with shifted arguments in
the non-linear term which take
into account the energy conservation. It can be reduced to the equation for the
moments \cite{DreminPhysLett} and solved. The most prominent feature of the
solution are
oscillations of cumulants (or ratio $H_q$) ; see \cite{DreminUFN}. It has
been confirmed by experiment as shown shown in fig. 1. It differs from all
previously considered phenomenological distributions and, from the mathematical
point of view, is interesting since it implies that QCD deals with
non-infinitely divisible distributions in contradistinction to conventional
ones. In particular, it forbids Poissonian cluster models so popular in
physics modelling.

It is interesting to note that it corresponds to rather smooth distribution,
i.e. there is no obvious one-to-one correspondence between the shapes of
distributions and behaviour of their cumulants.

\setcounter{equation}{0}
\def\theequation{4.\arabic{equation}}

\begin{center}
{\Large 4. APPLICATION TO PHOTONS.}
\end{center}

The above methods can easily be applied to photons \cite{Pavel}. Since the
generating functions of photon distributions for the squeezed and correlated
states are known \cite{obzor} one can use the equations of section 2
and get all the required moments. Let us consider the most general mixed
squeezed state of
one--mode light. The generating function of the photon number distribution was
obtained  in \cite{Ola1}:
\begin{eqnarray}
    G(u)=P_{0}\left[\left(1-\frac{u}{\lambda_{1}}\right)
    \left(1-\frac{u}{\lambda_{2}}\right)\right]^{-1/2}
    \mbox{exp}\left[\frac{u\xi_{1}}{u-\lambda_{1}}+\frac{u\xi_{2}}
    {u-\lambda_{2}}\right] \hspace{0.15cm} \mbox{,}
\end{eqnarray}
\begin{eqnarray}
  \lefteqn{}
    &&\lambda_{1}=\left(\sqrt{R_{11}R_{22}}
    -R_{12}\right)^{-1} \hspace{0.15cm}
    \mbox{,}
    \hspace{1cm} \lambda_{2}=-\left(\sqrt{R_{11}R_{22}}
    +R_{12}\right)^{-1}
    \hspace{0.15cm} \mbox{,}\nonumber\\
    &&\xi_{1}=
    \frac{1}{4}\left(1-\frac{R_{12}}{\sqrt{R_{11}R_{22}}}\right)
    \left(y_{1}^{2}R_{11}
    +y_{2}^{2}R_{22}-2\sqrt{R_{11}R_{22}}y_{1}y_{2}
    \right) \hspace{0.15cm} \mbox{,} \nonumber\\
    &&\xi_{2}
    =\frac{1}{4}\left(1+\frac{R_{12}}{\sqrt{R_{11}R_{22}}}\right)
    \left(y_{1}^{2}R_{11}+y_{2}^{2}R_{22}
    +2\sqrt{R_{11}R_{22}}y_{1}y_{2}
    \right) \hspace{0.15cm} \mbox{,} \nonumber
\end{eqnarray}
\begin {eqnarray}
  \lefteqn{}
    &&R_{11}= \left( T+2d+\frac{1}{2} \right) ^{-1}(\sigma _{pp}
    -\sigma _{qq}-2i\sigma _{pq}) =R_{22}^*,\nonumber\\
    &&R_{12}=\left( T+2d
     +\frac{1}{2} \right) ^{-1} \left( \frac{1}{2}-2d\right),
\end {eqnarray}
and
\begin{eqnarray}
  y_1={y_2}^*= \left( T-2d-\frac{1}{2} \right) ^{-1}
  \left[ (T-1) \langle z^* \rangle
  +(\sigma _{pp}-\sigma _{qq} +2i\sigma
  _{pq}) \langle z \rangle \right] .
\end{eqnarray}
The complex parameter $\langle z \rangle$ is given
by relation
\begin{eqnarray}
  \langle z \rangle =2^{-\frac{1}{2}} \left(
  \langle q \rangle +i \langle p \rangle \right) .
\label{defz}\end{eqnarray}
\begin {eqnarray}
  \lefteqn{}
    &&m_{11}=\sigma _{pp}
    =\mbox{Tr} \left( {\hat \varrho}{\hat p}^2 \right)
    -\langle p \rangle ^2,\nonumber\\
    &&m_{22}=\sigma _{qq}=\mbox{Tr} \left( {\hat
    \varrho}{\hat q}^2 \right) - \langle q \rangle ^2,\nonumber\\
    &&m_{12}=\sigma _{pq}=\frac{1}{2} \mbox{Tr} ~\left[
    {\hat \varrho}({\hat p\hat q}
    +{\hat q\hat p}) \right] -\langle p \rangle \langle q \rangle ,
\end{eqnarray}

  $$T=\mbox{Tr} \hspace{0.15cm} {\bf m} =\sigma _{pp}
  +\sigma _{qq}$$,
\noindent
  $$d=\det {\bf m} =\sigma _{pp}\sigma _{qq}-\sigma _{pq}^2$$,
and $\sigma _{ij}$ are the well known elements of the dispersion
matrix.

The photon distribution
function exhibits an oscillatory behaviour if we deal with highly
squeezed states (\hspace{0.05cm} $T=\sigma_{pp}
+\sigma_{xx}\gg 1$ \hspace{0.05cm} )
for large values of the parameter $z$.
A question arises: is it possible to obtain a similar ``abnormal''
behaviour of other characteristics of the photon distribution, namely,
introduced in section 2
cumulants, factorial moments and their ratio $H_{q}$ ?
If yes, then in what region of parameters can such
anomalies reveal themselves?

The direct differentiation of function $\mbox{ln}~G(u)$ at $u=1$ yields
the cumulants (see section 2)
\begin{eqnarray}
    K_{q}=\frac{(q-1)!}{\langle n \rangle^{q}}\left[\frac{1}
    {(\lambda_{1}-1)^{q}}
    \left(\frac{1}{2}+q\frac{\xi_{1}\lambda_{1}}{1-\lambda_{1}}\right)
    +\frac{1}{(\lambda_{2}-1)^{q}}
    \left(\frac{1}{2}+q\frac{\xi_{2}\lambda_{2}}{1-\lambda_{2}}\right)
    \right],
\label{gencum}\end{eqnarray}
with the average number of photons $\langle n \rangle$  \cite{Ola1}
\begin{eqnarray}
    \langle n \rangle=\frac{T-1}{2}+|z|^{2} \hspace{0.15cm}
    \mbox{.} \nonumber
\end{eqnarray}
Now let us consider the case of the slightly squeezed state,
$y=(T-1)\ll 1$ when the photon distribution function does not
oscillate. Impose also an additional condition
\begin{eqnarray}
    \gamma=\frac{|z|^{2}}{\sqrt{y/2}}\gg 1
    \hspace{0.15cm} \mbox{,} \nonumber
\end{eqnarray}
that makes possible to obtain approximate formulas for the
functions $K_{q}$, $F_{q}$, and $H_{q}$.
For $K_{q}$, we have the following approximate expression:
\begin{eqnarray}
    K_{q}=q!(-1)^{q-1}\gamma^{1-q} \hspace{0.15cm} \mbox{.}
\end{eqnarray}
Then recursion relation (2.9) yields:
\begin{eqnarray}
    F_{q}=q!(-1)^{q}\gamma^{-q}L_{q}^{-1}(\gamma)
    \hspace{0.15cm} \mbox{,}
\end{eqnarray}
where $L_{q}^{-1}(x)$ are generalized Laguerre polynomials.
For $H_{q}$, with $ q \ll \gamma $ we have:
\begin{eqnarray}
    H_{q}=K_{q}/F_{q}
     =-\frac{\gamma}{L_{q}^{-1}(\gamma)}\approx (-1)^{q+1}
    q!\gamma^{1-q} \ll 1 \hspace{0.15cm} \mbox{.}
\end{eqnarray}
(If $\gamma\gg q$ the term with highest power of $\gamma$ dominates
over the rest of sum in $L_{q}^{-1}(\gamma)$, and
$F_{q} \rightarrow 1$ as for Poisson distribution). The exact shape
of the function $H_{q}$ is shown in fig.$2^{a}$.
The distribution function $P_{n}$ does not oscillate (fig.$2^{b}$).

However, the most abrupt oscillations of the functions $K_{q}$ and
$H_{q}$ have been obtained when $(T-1)\ll 1$, but condition
$\gamma\gg 1$ is not valid. The corresponding curves are shown in
figs.$3^{a}$, $3^{b}$. Note that the photon distribution function is
smooth again being approximately equal to zero at $q\neq 1$.

The most regular oscillating patterns of $K_{q}$ and $H_{q}$ are
seen at $(T-1)\sim 0.1$, $|z|\sim 1$ (figs.$4^{a}$, $4^{b}$).
Finally we consider the opposite case when the photon
distribution function
$P_{n}$ exhibits strong oscillations while $K_{q}$ and $H_{q}$
behave smoothly. Such a behaviour is typical at $T\sim 100$,
$|z|\sim 1$  when $K_{q}$ exponentially grows while
$H_{q}$ monotonically decreases with $q$ (fig.$5$).

Thus we have shown that the cumulants of the photon distribution
function for one--mode squeezed and correlated light at finite
temperature possess strongly oscillating behaviour in the region of
slight squeezing where the photon distribution function itself has
no oscillations. And vice versa in the region of large squeezing,
where the photon distribution function strongly oscillates, the
cumulants behave smoothly. Hence, the behaviour of cumulants may
provide a very sensitive method of detecting very small
squeezing and correlation phenomena due to the presence of strong
oscillations.

Let us note also that these methods were successfully applied to
predicting some regularities in behaviour of lasers in the lasing
 regime \cite{hwa}.
\newpage
\setcounter{equation}{0}
\def\theequation{5.\arabic{equation}}

\begin{center}
{\Large 5. COHERENT AND SQUEEZED STATES IN PION PHYSICS.}
\end{center}

The notion of coherent and squeezed states is not widely applied in pion
production yet even though several papers deal with the subject \cite{hs},
\cite{bss},\cite{andr},\cite{wein},\cite{bj},\cite{ans},\cite{kowt},\cite{rawi},\cite{kogan},\cite{gamu}. The most striking difference with photon case
appears due to the requirement of isotopic spin conservation inherent in
pion physics but absent for photons. It gives rise to strong charge asymmetry
of produced events. If the isospin zero projection of the coherent state
is considered, it predicts many events without
neutral pions, for example. It is demonstrated in fig. 6 borrowed from
\cite{andr}. It used to be related to the so-called Centauro-type events
observed in cosmic rays.

Let us consider
for definitiveness the process of the collision of two high energy
nucleons, where in the final state, apart from the two nucleons,
 a certain number of pions has been created. The total isotopic spin of the
pion
system is limited, according to the conservation laws, to the values
$I=0,1,2$,  whereas in the general case it could take values up to
$n_{tot}$, where $n_{tot}$ is a total number of pions. This fact
significantly affects the pion charge distribution. Even if the
distribution in the total pion number $n_{tot}$ is poissonian (as it happens
when pions are produced by classical currents), it turns
out, that the separate distributions of the charged  ($n_{+}, n_{-}$) or
neutral
($n_0$) pions are much wider than the Poisson one (see fig.6).  The
experimental confirmation of this fact
could be the "Centauro" events, where the charged particles are noticeably
dominating, or "anti-Centauro" ones with a large number of neutral
pions.

Let us illustrate the idea by considering, following the
quasiclassical approach of refs.\cite{andr},\cite{ans}, the production of many
pions in
a system with zero isotopic spin. The characteristic initial assumption
is a possibility of describing the pion system that radiates the
final state pions as a classical field , i.e. that the number
of pions per phase space cell is assumed to be big. According to the
standard reduction formula, the amplitude of generation of $N$ pions
by the source $J$ equals
\begin{eqnarray}
A^{a_{1},\ldots,a_{n}} (k_1,\ldots,k_n) =
{\lim_{k^2_n \rightarrow m^2_{\pi}}} \int D\pi^a \int DJ^a W[J] \nonumber \\
\exp (iS[\pi]+i\int d^4x \pi^a J^a ) \prod_{n=1}^N \int d^4 x_n
e^{ik_n x_n} (-\partial^2_{x_n} - m^2_{\pi}) \pi^{a_n} (x_n),
\end{eqnarray}
where the functional integration over $J$ corresponds to the averaging
over the characteristics of the pion source. Let us notice, that the
radiation of a classical current exactly reproduces the language of
coherent states. The specific feature of pion fields is that one
projects the coherent states onto the states with a definite isotopic
spin, and it gives rise to drastical change of final charge distributions.
 The quasiclassical estimate of the amplitude, performed
in the assumption on the axial symmetry of the initial interaction and
on the isotopic symmetry of the pion system (i.e. of the zero total
isospin) \cite{bss},\cite{andr},\cite{ans}, leads to the two characteristic
conclusions that also
appear in other publications on this topic.

Firstly, only the distribution over the total number of pions is
poissonian. The distributions over the number of neutral and
charged pions are much wider (see fig. 6). For the state with zero isospin
a probability of finding $2n$ neutral pions in the system of $2N$
pions has a form \cite{hs},\cite{kowt}:
\begin{equation}
P(n,N) = {(N!)^2 2^{2n} (2n)! \over (n!)^2 2^{2N} (2N+1)!} .\label{PnN}
\end{equation}
At large $n,N$ we have a characteristic distribution \cite{ansry}:
\begin{equation}
P(n,N) \sim (n/N)^{-1/2} .\label{f}
\end{equation}
Secondly, the conservation of the total isospin leads, for example,
to the specific angular correlations between the particles with different
charges. Let us give the characteristic formula for the correlation over
the azimuthal angle $\varphi$, obtained in \cite{bldi} for the pions having
zero rapidity:
\begin{equation}
{(\sigma_{tot} {d\sigma^{\pi^+ \pi^-} \over dk_1 dk_2}
-{9 \over 10} {d\sigma ^{\pi^+} \over dk_1} {d\sigma ^{\pi^-} \over dk_2})
\over ({d\sigma ^{\pi^+} \over dk_1} {d\sigma^{\pi^-} \over dk_2})}
 ={3 \over 10} \cos ^2 (\varphi_1 - \varphi_2) .
 \end{equation}
The experimental verification of such predictions is to our opinion very
interesting.

The similar procedure of projecting {\it squeezed} states onto the states with
definite isospin has been attempted in \cite{drh}. It produced some interesting
modifications of fig. 6 which we have no space to discuss here.

In the recent literature a "disoriented chiral condensate" was widely
discussed as a possible asymmetry source in the production of charged
and neutral pions in some fraction of the events ( in particular, in
the above-mentioned "Centauros"). Let us remind, that the
transformation properties of mesons with respect to the chiral
transformations are determined according  to the corresponding properties
of the order parameter, which characterizes the spontaneous breaking
of chiral symmetry and is given, for example, by the average from the
bilinear combination of quark fields
\begin{equation}
\Phi \sim \langle {\bar{q}}_L q_R \rangle,\label{Phi}
\end{equation}
where $q_{R(L)}$ are the right (left) states of the massless quarks. For
investigating the character of the singularity of thermodynamical functions
in the vicinity of the phase transitions, it is desirable to find a solvable
model having the same symmetry. Then, according to the universality principle
based on the scale invariance near the critical point, the solutions of this
model will have the same set of singularities. In such an approach the order
parameter (\ref{Phi}) can be rewritten in terms of a set of hadron fields
having
the same symmetry. Therefore, the multipion states become related to the
massless
quark fields and quark-gluon plasma, i.e. there appears a hope to
describe a phase transition between these so differing  phases within such a
model. In the realistic case of two massless quark flavors the chiral field
can be written in the form
\begin{equation}
\Phi \sim \sigma \bullet {\bar{1}} + i {\bar{\tau}}
\bullet {\bar{\pi}},
\end{equation}
where $\sigma, {\bar{\pi}}$ are the real fields, ${\bar{\tau}}$ are the
standard Pauli matrices, the $\pi$ - meson fields ${\bar{\pi}}$ form
an isotriplet and $\sigma$ is an isosinglet.

In the case of $SU(3)$ - algebra the number of such fields is already $18$.
They form the scalar and pseudoscalar nonets.

 The dynamics of these degrees
of freedom can be described by the lagrangian of the linear $\sigma$ model
\begin{equation}
{\it{L}}   = {1 \over 2} (\partial \sigma)^2+
{1 \over 2} (\partial {\bar{\pi}})^2-V(\sigma,{\bar{\pi}}) ,\end{equation}
where $V$ is a potential depending on the combination $\sigma^2+{\bar{\pi}}^2$.
In the standard version \cite{GML} the spontaneous symmetry breaking occurs via
the formation of the nonzero vacuum average of the field $\sigma$. The
isotriplet
fields remain massless, i.e. the pions are the goldstones of the chiral
group.

Let us now assume \cite{bj}, that in some region of space the vacuum
orientation
is different from the standard one, and, for example,
\begin{equation}
\langle \sigma \rangle = f_{\pi} \cos \theta, \;\;\;\;
\langle {\bar{\pi}} \rangle = f_{\pi} {\bar{n}} \sin \theta,
\end{equation}
 where $f_{\pi}=$ 93 MeV, and ${\bar{n}}$ is a unit orientation vector of
 ${\bar{\pi}}$.  Such an assumption presupposes a specific scenario of
the process, which is still not studied in details.

If the field $\Phi$ is isotropic with respect to a direction on the
3-dimensional sphere in the 4-dimensional space with the angles defined as
\begin{equation}
(\sigma,\pi_3,\pi_2,\pi_1)=
(\cos \theta, \sin \theta \cos \phi, \sin \theta \sin \phi \sin \eta,
\sin \theta \sin \phi \cos \eta),
\end{equation}
then for the probability distribution of a given state $r \equiv \cos ^2 \phi$
we have:
\begin{equation}
\int_{r_1}^{r_2} dr P(r)=
{1 \over \pi^2} \int_{0}^{2 \pi} d \eta \int_0^{\pi}  d \theta \sin ^2 \theta
\int_{\arccos r_2^{1/2}}^{\arccos r_1^{1/2}} d \phi \sin \phi,
\end{equation}
as it was obtained for the first time in refs.\cite{andr,ansry}:
\begin{equation}
P(r)={1 \over 2 {\sqrt{r}}},
\end{equation}
i.e. we have again returned to the formula (\ref{PnN}). Thus the probability
of finding an event with a fraction of the neutral pions being smaller
than a certain given $r_0$ is
\begin{equation}
W(r<r_0) = {\sqrt{r_0}} ,
\end{equation}
which constitutes 10 \% even at $r_0$ = 0.01. The charge fluctuations in
such a system are much larger than those that would follow from the Poisson
distribution, where the distributions are concentrated around $r$=1/3.

Let us note that the model of creation of $N$ isoscalar pion pairs is
described as
\begin{equation}
|\Psi \rangle \propto (2a_{+}^{+}a_{-}^{+}-(a_{0}^{+})^2)^N|0\rangle ,
\end{equation}
which is reminiscent of squeezed states projected onto zero isospin but
is more restrictive than that projection. (Here $a_{+,-,0}^{+}$ are the
creation operators of positive, negative, neutral pions, correspondingly).

\setcounter{equation}{0}
\def\theequation{6.\arabic{equation}}

\begin{center}
{\Large 6. CONCLUSIONS}
\end{center}

We have shown that the correlation analysis can be done by similar methods for
photons and pions. Some further understanding of the relation between
distribution behaviour and features of its moments is awaited for. More clear
insight into usefulness of the idea of coherent and squeezed states as applied
to particle physics is required. However, the very first attempts in these
directions described above are very encouraging.


\newpage
\begin{thebibliography}{99}
\bibitem{UA5}
UA5 coll.,  Alner G.J. et al., Phys. Rep. 154 (1987) 247.
\bibitem{DELPHI}
DELPHI coll., Abreu P. et al., Z. Phys. C50 (1991) 185.
\bibitem{OPAL}
OPAL coll., Acton P.D. et al., Z. Phys. C53 (1992) 539.
\bibitem{Dremin} Dremin I.M., Mod. Phys. Lett. A 8 (1993) 2747.
\bibitem{DreminPhysLett} Dremin I.M., Phys. Lett. B 313 (1993) 209; \\
                   Dremin I.M. and Hwa R.C., Phys. Rev. D 49 (1994) 5805.
\bibitem{DreminUFN} Dremin I.M., Usp. Fiz. Nauk 164 (1994) 785;
                    Physics --Uspekhi, 37 (1994) 715.
\bibitem{Gian} Dremin I.M., Arena V., Boca G., et al., Phys. Lett. B 336
               (1994) 119.
\bibitem{Pavel} Dodonov V.V., Dremin I.M., Polynkin P.G., and Man'ko V.I.,
                Phys. Lett. A 193 (1994) 209.
\bibitem{5}
Dokshitzer Yu.L., Khoze V.A., Mueller A.H. and Troyan S.I., Basics of
perturbative QCD, Gif-sur-Yvette, Editions Frontieres, 1991.

\bibitem{obzor} Dodonov V.V., Dremin I.M., Polynkin P.G., Man'ko O.V.,
and Man'ko V.I.,  hep-ph 9502394.
\bibitem{Ola1} Dodonov V.V., Man'ko O.V., and Man'ko V.I.,
               Phys. Rev. A 49 (1994) 2993.
\bibitem{hwa} Hwa R.C., Phys. Rev. C 50 (1994) 383.
\bibitem{hs} Horn~D., Silver~R., Ann. Phys. (N.Y.) 66
  (1971) 509.
\bibitem{bss} Botke~J.C., Scalapino~D.J., Sugar~R.L., Phys. Rev.
 D9 (1974) 813; D10 (1974) 1604.
\bibitem{andr} Andreev~I.V., Pisma  v ZhETF 33 (1981) 384 (JETP Lett. 33 (1981)
367).
\bibitem{wein} Vourdas A., and Weiner R.M., Phys. Rev. A 36
                      (1987) 5866.
\bibitem{bj} Bjorken~J.D., Acta Physica Polonica B23
  (1992) 561. \\ Bjorken~J.D., Kowalski~K.L., Taylor~C.C.,
SLAC-PUB-6109 (1993).
\bibitem{ans} Anselm~A.A., Phys. Lett. B217 (1989) 169.
\bibitem{kowt} Kowalski~K.L., Taylor~C.C., CWRUTH-92-6, hep-ph/9211282 (1992).
\bibitem{rawi} Rajagopal~K., Wilczek~F., Nucl. Phys. B379
  (1993) 395; {\it{ibid}} B404 (1993) 577.
\bibitem{kogan} Kogan~I.I., Phys. Rev. D48 (1993) 3971.
\bibitem{gamu} Gavin~S., M\"uller~B., Phys. Lett. B329 (1994) 486.
\bibitem{ansry} Anselm~A.A., Ryskin~M.G., Phys. Lett. B266
  (1991) 482.

\bibitem{bldi} Blaizot~J.P., Krzywicki~A., Phys. Rev. D46
  (1992) 246.

\bibitem{drh} Dremin I.M., and Hwa R.C. (in preparation).
\bibitem{GML} Gell-Mann~M., Levy~M., Nuovo Cimento 16
  (1960) 705.

\end{thebibliography}
\end{document}